\documentclass[prd,nofootinbib,preprint,superscriptaddress]{revtex4}

\usepackage{tikz,xcolor,hyperref}\usepackage{amsmath, amssymb, amsthm, graphicx, epsfig, fancyhdr,epsfig, slashed}

\usepackage{tikzsymbols}
\usepackage{natbib}
\usepackage{float}

\newtheorem{theorem}{Theorem}
\usepackage{float}
\usepackage{bm}
\usepackage{url}

\newcommand{\be}{\begin{equation}}
\newcommand{\ee}{\end{equation}}
\newcommand{\bea}{\begin{eqnarray}}
\newcommand{\eea}{\end{eqnarray}}

\definecolor{lime}{HTML}{A6CE39}
\DeclareRobustCommand{\orcidicon}{
	\begin{tikzpicture}
	\draw[lime, fill=lime] (0,0) 
	circle [radius=0.2] 
	node[white] {{\fontfamily{qag}\selectfont \tiny ID}};
	\draw[white, fill=white] (-0.0625,0.095) 
	circle [radius=0.007];
	\end{tikzpicture}
	\hspace{-2mm}
}

\foreach \x in {A, ..., Z}{\expandafter\xdef\csname orcid\x\endcsname{\noexpand\href{https://orcid.org/\csname orcidauthor\x\endcsname}
			{\noexpand\orcidicon}}
}

\begin{document}


\title{Casimir effect in Yang-Mills theories}


\author{Marco Frasca\orcidA{}}\email{marcofrasca@mclink.it}
\affiliation{via Erasmo Gattamelata, 3, 00176 Roma, Italy}


\date{\today}

\begin{abstract}
We study both massless scalar and Yang-Mills field theories in the deep infrared in presence of a simple boundary. We can show, with the help of the recent scenario emerging from studies on the propagators, that the presence of a mass gap makes the Casimir contribution exponentially small as should be expected. 
We obtain our result from the solution of the Dyson-Schwinger set of equations obtained in form of partial differential equations. Our results agree fairly well with recent lattice computations.
\end{abstract}


\maketitle

\section{Introduction}

Vacuum in quantum field theory is known to not be inert. The most striking evidence of this was put forward on 1945 by Casimir \cite{Casimir:1948dh} providing an example of a macroscopic effect produced by quantum fluctuations. Due to the smallness of the involved elements, an experimental proof of the Casimir effect has been obtained quite recently (see \cite{Bordag:2001qi} for a review).

Casimir effect is an example of a non-trivial behavior of a quantum field theory due to the presence of boundary conditions that somehow modify the behavior of the free space case. This kind of question is interesting {\sl per se}. But while in the ultraviolet regime we have techniques able to cope with this kind of problems, in the opposite limit, the infrared regime, such techniques were generally lacking. It is important to note that the behavior at different energy regimes of a quantum field theory could result quite different depending on the structure of vacuum. So, we know that at high energies, QCD behaves as an almost free theory and asymptotic states of the Yang-Mills field are massless gluons. On the other side, at lower energies, a Yang-Mills theory displays a mass gap yielding a very different behavior in this case. E.g. this could be inherent to a non-trivial non-perturbative vacuum seen as an instanton liquid \cite{Schafer:1996wv}.

Studies of quantum field theories in the infrared limit have had an important rebirth in this last two decades by the use of Dyson-Schwinger equations \cite{von Smekal:1997is,von Smekal:1997vx} and by the improvement in computing resources that made possible to analyze Yang-Mills theory on larger lattices. The theoretical proposal to solve the infinite hierarchy of Dyson-Schwinger equations was a truncation scheme that, for the Landau gauge, produced a gluon propagator going to zero and a ghost propagator running to infinity faster than the free case with momenta going to zero. Such a scenario was previously devised by Daniel Zwanziger \cite{Zwanziger:1991gz} providing also a criterion for color confinement then dubbed Gribov-Zwanziger criterion. Initial lattice computations seemed to support these conclusions even if no bending toward zero of the gluon propagator was ever observed at lower momenta. People generally thought that was just a matter of volumes and, increasing computational resources, things should have changed.

The breakthrough come out on 2007. At the Lattice 2007 Conference in Regensburg, three groups presented their results with huge volumes arriving to such a significant value as $(27fm)^4$ \cite{Bogolubsky:2007ud,Cucchieri:2007md,Oliveira:2007px}. The shocking result was that the scenario devised since then, generally accepted as correct, was not describing the situation seen on the lattice: The gluon propagator was reaching a plateau at lower momenta with a finite non-zero value in zero and the ghost propagator was behaving as that of a free massless particle. All in all, the running coupling was seen to bend clearly toward zero without evidence of a non-trivial fixed point as was expected instead.

In the eighties, a classical paper by Cornwall \cite{Cornwall:1981zr} showed that indeed the gluon propagator showed a dynamical mass, dependent on momenta, that in the zero limit reaches a finite constant value. With the emerging of techniques to solve Dyson-Schwinger equations in the nineties, it was straightforward to try to solve them numerically. This numerical approaches showed that Cornwall view is indeed correct \cite{Aguilar:2004sw} but this paper displayed also all the scenario seen since 2007 on lattice computations. 
So, a research line developed producing an in-depth theoretical and numerical analysis of Dyson-Schwinger equations supporting the current view \cite{Aguilar:2006gr,Aguilar:2008xm,Aguilar:2011ux,Aguilar:2011yb,Binosi:2009qm,RodriguezQuintero:2011vw,RodriguezQuintero:2010wy,Boucaud:2010gr,Boucaud:2008gn,Boucaud:2008ji}.

The idea in this paper is to start from this situation, also supported by theoretical arguments, to analyze the behavior of the vacuum in the infrared limit. The best way to see a non-trivial behavior is through the analysis of the deep infrared limit with simple boundary conditions. As we will see, this can be accomplished yielding the result, somewhat expected with a mass gap, that the Casimir contribution is exponentially damped both for scalar and Yang-Mills theories. The approach we will use to evaluate the Casimir effect is due to Schwinger \cite{schw1,schw2}.

The paper is so structured. In sec.\ref{IL} we introduce the mapping theorem between a scalar field theory and a non-Abelian gauge field theory. In sec.\ref{CL} we derive the 1P- and 2P-correlation functions for the Yang-Mills theory solving the Dyson-Schwinger set of equations written as partial differential equations (PDE). In sec.\ref{EF} we present our main results on the Casimir contribution. Finally, in sec.\ref{conc} we give our conclusions.

\section{Mapping theorem}
\label{IL}

Our aim is to solve the Dyson-Schwinger set of equations for the Yang-Mills theory to evaluate the corresponding effective action. This question can be approached through a mapping theorem proved recently \cite{Frasca:2007uz,Frasca:2009yp} that can be stated in the following form:

\begin{theorem}[Mapping]
\label{teo1}
An extremum of the action
\begin{equation}
    S = \int d^4x\left[\frac{1}{2}(\partial\phi)^2-\frac{\lambda}{4}\phi^4\right]
\end{equation}
is also an extremum of the SU(N) Yang-Mills Lagrangian when one properly chooses $A_\mu^a$ with some components being zero and all others being equal, and $\lambda=Ng^2$, being $g$ the coupling constant of the Yang-Mills field, when only time dependence is retained. In the most general case the following mapping holds
\begin{equation}
    A_\mu^a(x)=\eta_\mu^a\phi(x)+O(1/\sqrt{N}g)
\end{equation}
being $\eta_\mu^a$ constant, that becomes exact for the Lorenz gauge.
\end{theorem}

The proof of this theorem was completed in Ref.\cite{Frasca:2009yp} to answer a criticism by Terence Tao. Tao finally agreed with this conclusion\cite{Tao}. Stated otherwise, this theorem determines an asymptotic mapping between the solutions of the two classical theories when the couplings are taken large enough. An incomplete form of this mapping was already stated in Ref.\cite{Smilga:2001ck}. The existence of this mapping grants that we can write down the propagator of Yang-Mills theory, e.g. in the Landau gauge, as
\begin{equation}
\label{eq:ymprop}   
  D_{\mu\nu}^{ab}(p)=\delta_{ab}\left(\eta_{\mu\nu}
  -\frac{p_\mu  p_\nu}{p^2}\right)\Delta(p),
\end{equation}
with $\Delta(p)$ being the same propagator of the scalar field theory.  

\section{Dyson-Schwinger method}
\label{CL}

We derive the Dyson-Schwinger equations for the Yang-Mills field for the 1P- and 2P-functions, using the Bender-Milton-Savage technique \cite{Bender:1999ek}.

The correlation functions are obtained when a given exact solution is known for the one-point function i.e., one has to solve exactly the equations
\begin{equation}
\partial^\mu\partial_\mu A^a_\nu-\frac{1}{2\alpha}\partial_\nu(\partial^\mu A^a_\mu)+gf^{abc}A^{b\mu}(\partial_\mu A^c_\nu-\partial_\nu A^c_\mu)+gf^{abc}\partial^\mu(A^b_\mu A^c_\nu)+g^2f^{abc}f^{cde}A^{b\mu}A^d_\mu A^e_\nu = 0.
\end{equation} 
In the Landau gauge ($\alpha\rightarrow 0$), these are exactly given in the form
\begin{equation}
\label{eq:exact}
   A^a_\nu(x)=\eta^a_\nu\left(\frac{2}{Ng^2}\right)^\frac{1}{4}\mu\cdot{\rm sn}(px,-1),
\end{equation}
with ${\rm sn}(px,-1)$ the Jacobi snoidal elliptic function and $\eta_\mu^a$ being a set of constants to be determined depending on the problem at hand (e.g., for $SU(2)$ one can take $\eta_1^1=\eta_2^2=\eta_3^3=1$, all other components being zero) and $\mu$ an integration constant with the dimension of an energy. This holds provided the following dispersion relation holds
\begin{equation}
\label{eq:disp}
    p^2=\sqrt{\frac{Ng^2}{2}}\mu^2.
\end{equation}
Solutions given in eq.(\ref{eq:exact}) appear as massive solution, due to the dispersion relation (\ref{eq:disp}), even if we started from a massless theory. 

Then, if we use these solutions as one-point function of the set of Schwinger--Dyson equations for a non-Abelian gauge theory without fermions, given by \cite{Frasca:2015yva}, we are able to compute the two-point functions exactly, without any approximation or truncation. We use the approach devised in \cite{Bender:1999ek}. Indeed, to get the Schwinger--Dyson equations one has to start from the quantum equations of motion that have the form
\begin{eqnarray}
   &&\partial^\mu\partial_\mu A^a_\nu+gf^{abc}A^{b\mu}(\partial_\mu A^c_\nu-\partial_\nu A^c_\mu)+gf^{abc}\partial^\mu(A^b_\mu A^c_\nu)+g^2f^{abc}f^{cde}A^{b\mu}A^d_\mu A^e_\nu \nonumber \\
	&&= gf^{abc}\partial_\nu(\bar c^b c^c) + j_\nu^a, \nonumber \\
	 &&\partial^\mu\partial_\mu c^a+gf^{abc}\partial^\mu(A_\mu^bc^c)=\varepsilon^a. 
\end{eqnarray}
We fix the gauge to the Landau gauge, $\alpha\rightarrow 0$, and $c,\ \bar c$ are the ghost fields. Averaging on the vacuum state and dividing by the partition function $Z_{YM}[j,\bar\varepsilon,\varepsilon]$, one has
\begin{eqnarray}
    &&\partial^2G_{1\nu}^{(j)a}(x)+
		gf^{abc}(\langle A^{b\mu}\partial_\mu A^c_\nu\rangle-\langle A^{b\mu}\partial_\nu A^c_\mu\rangle)Z^{-1}_{YM}[j,\bar\varepsilon,\epsilon]
		+gf^{abc}\partial^\mu\langle A^b_\mu A^c_\nu\rangle Z^{-1}_{YM}[j,\bar\varepsilon,\varepsilon]
		\nonumber \\
		&&+g^2f^{abc}f^{cde}\langle A^{b\mu}A^d_\mu A^e_\nu\rangle Z^{-1}_{YM}[j,\bar\varepsilon,\varepsilon] 
		=gf^{abc}\langle\partial_\nu(\bar c^b c^c)\rangle Z^{-1}_{YM}[j,\bar\varepsilon,\varepsilon] + j_\nu^a, \nonumber \\
	 &&\partial^2 P^{(\varepsilon)a}_1(x)
	 +gf^{abc}\partial^\mu\langle A_\mu^bc^c\rangle Z^{-1}_{YM}[j,\bar\varepsilon,\varepsilon]=\varepsilon^a. 
\end{eqnarray}
The one-point function is given by
\begin{eqnarray}
    &&G_{1\nu}^{(j)a}(x)Z_{YM}[j,\bar\varepsilon,\epsilon]=\langle A^a_\nu(x)\rangle, \nonumber \\
		&&P^{(\varepsilon)a}_1(x)Z_{YM}[j,\bar\varepsilon,\epsilon]=\langle c^a(x)\rangle. 
\end{eqnarray}
Deriving once with respect to currents, at the same point because of the averages on the vacuum (see \cite{Bender:1999ek}), one has
\begin{eqnarray}
   &&G_{2\nu\kappa}^{(j)ab}(x,x)Z_{YM}[j,\bar\varepsilon,\epsilon]+G_{1\nu}^{(j)a}(x)G_{1\kappa}^{(j)b}(x)Z_{YM}[j,\bar\varepsilon,\epsilon]=\langle A^a_\nu(x)A^b_\kappa(x)\rangle, \nonumber \\
	 &&P^{(\varepsilon)ab}_2(x,x)Z_{YM}[j,\bar\varepsilon,\epsilon]+\bar P^{(\varepsilon)a}_1(x)P^{(\varepsilon)b}_1(x)Z_{YM}[j,\bar\varepsilon,\epsilon]=
	\langle \bar c^b(x)c^a(x)\rangle, \nonumber \\
	&&\partial_\mu G_{2\nu\kappa}^{(j)ab}(x,x)Z_{YM}[j,\bar\varepsilon,\epsilon]+\partial_\mu G_{1\nu}^{(j)a}(x)G_{1\kappa}^{(j)b}(x)Z_{YM}[j,\bar\varepsilon,\epsilon]=
	\langle\partial_\mu A^a_\nu(x)A^b_\kappa(x)\rangle, \nonumber \\
	&&K^{(\varepsilon,j)ab}_{2\nu}(x,x)Z_{YM}[j,\bar\varepsilon,\epsilon]+P^{(\varepsilon)a}_1(x)G_{1\nu}^{(j)b}(x)Z_{YM}[j,\bar\varepsilon,\epsilon]=\langle c^a(x)A_\nu^b(x)\rangle,
\end{eqnarray}
and twice
\begin{eqnarray}
   &&G_{3\nu\kappa\rho}^{(j)abc}(x,x,x)Z_{YM}[j,\bar\varepsilon,\epsilon]+G_{2\nu\kappa}^{(j)ab}(x,x)G_{1\rho}^{(j)c}(x)Z_{YM}[j,\bar\varepsilon,\epsilon]+ \nonumber \\
	&&G_{2\nu\rho}^{(j)ac}(x,x)G_{1\kappa}^{(j)b}(x)Z_{YM}[j,\bar\varepsilon,\epsilon]
	+G_{1\nu}^{(j)a}(x)G_{2\kappa\rho}^{(j)bc}(x,x)Z_{YM}[j,\bar\varepsilon,\epsilon]+ \nonumber \\
	&&G_{1\nu}^{(j)a}(x)G_{1\kappa}^{(j)b}(x)G_{1\rho}^{(j)c}(x)Z_{YM}[j,\bar\varepsilon,\epsilon]=\langle A^a_\nu(x)A^b_\kappa(x)A^c_\rho(x)\rangle.
\end{eqnarray}
These give us the first set of Schwinger--Dyson equations as
\begin{eqnarray}
\label{eq:ds_1}
    &&\partial^2G_{1\nu}^{(j)a}(x)+gf^{abc}(
		\partial^\mu G_{2\mu\nu}^{(j)bc}(x,x)+\partial^\mu G_{1\mu}^{(j)b}(x)G_{1\nu}^{(j)c}(x)-
		\partial_\nu G_{2\mu}^{(j)\mu bc}(x,x)-\partial_\nu G_{1\mu}^{(j)b}(x)G_{1}^{(j)\mu c}(x)) \nonumber \\
		&&+gf^{abc}\partial^\mu G_{2\mu\nu}^{(j)bc}(x,x)+gf^{abc}\partial^\mu(G_{1\mu}^{(j)b}(x)G_{1\nu}^{(j)c}(x))		
		\nonumber \\
		&&+g^2f^{abc}f^{cde}(G_{3\mu\nu}^{(j)\mu bde}(x,x,x)
		+G_{2\mu\nu}^{(j)bd}(x,x)G_{1}^{(j)\mu e}(x) \nonumber \\
	&&+G_{2\nu\rho}^{(j)eb}(x,x)G_{1}^{(j)\rho d}(x)
	+G_{2\mu\nu}^{(j)de}(x,x)G_{1}^{(j)\mu b}(x)+ \nonumber \\
	&&G_{1}^{(j)\mu b}(x)G_{1\mu}^{(j)d}(x)G_{1\nu}^{(j)e}(x))
		=gf^{abc}(\partial_\nu P^{(\varepsilon)bc}_2(x,x)+\partial_\nu (\bar P^{(\varepsilon)b}_1(x)P^{(\varepsilon)c}_1(x))) 
		+ j_\nu^a, \nonumber \\
	 &&\partial^2 P^{(\varepsilon)a}_1(x)+gf^{abc}\partial^\mu
	(K^{(\varepsilon,j)bc}_{2\mu}(x,x)+P^{(\varepsilon)b}_1(x)G_{1\mu}^{(j)c}(x))=\varepsilon^a. 
\end{eqnarray}
By setting the currents to zero and noticing that, by translation invariance, one has $G_2(x,x)=G_2(x-x)=G_2(0)$, $G_3(x,x,x)=G_3(0,0)$ and $K_2(x,x)=K_2(0)$, we get
\begin{eqnarray}
    &&\partial^2G_{1\nu}^{a}(x)+gf^{abc}(
		\partial^\mu G_{2\mu\nu}^{bc}(0)+\partial^\mu G_{1\mu}^{b}(x)G_{1\nu}^{c}(x)-
		\partial_\nu G_{2\mu}^{\nu bc}(0)-\partial_\nu G_{1\mu}^{b}(x)G_{1}^{\mu c}(x)) \nonumber \\
		&&+gf^{abc}\partial^\mu G_{2\mu\nu}^{bc}(0)+gf^{abc}\partial^\mu(G_{1\mu}^{b}(x)G_{1\nu}^{c}(x))		
		\nonumber \\
		&&+g^2f^{abc}f^{cde}(G_{3\mu\nu}^{\mu bde}(0,0)
		+G_{2\mu\nu}^{bd}(0)G_{1}^{\mu e}(x) \nonumber \\
	&&+G_{2\nu\rho}^{eb}(0)G_{1}^{\rho d}(x)
	+G_{2\mu\nu}^{de}(0)G_{1}^{\mu b}(x)+ \nonumber \\
	&&G_{1}^{\mu b}(x)G_{1\mu}^{d}(x)G_{1\nu}^{e}(x))
		=gf^{abc}(\partial_\nu P^{bc}_2(0)+\partial_\nu (\bar P^{b}_1(x)P^{c}_1(x))), \nonumber \\
	 &&\partial^2 P^{a}_1(x)+gf^{abc}\partial^\mu
	(K^{bc}_{2\mu}(0)+P^{b}_1(x)G_{1\mu}^{c}(x))=0. 
\end{eqnarray}
This set of Schwinger--Dyson equations can be solved exactly in the Landau gauge with the aforementioned exact solutions. 
This is so by noting that the contributions coming from $G_{2\mu\nu}^{ab}(0)$, $P_2^{ab}(0)$, $G_{3\mu\nu}^{\mu bde}(0,0)$ and $K^{bc}_{2\mu}(0)$ are zero in this case due to the fact that they give a symmetric group contribution against the antisymmetric structure constants of the group itself. Then, one gets that the ghost one-point function decouples and can be assumed to be a constant and does not contribute to the gluon one-point function.

The Schwinger--Dyson equation for the two-point functions can be obtained by further deriving eq. (\ref{eq:ds_1}). One has
\begin{eqnarray}
\label{eq:ds_2}
    &&\partial^2G_{2\nu\kappa}^{(j)am}(x-y)
		+gf^{abc}(
		\partial^\mu G_{3\mu\nu\kappa}^{(j)bcm}(x,x,y)
		+\partial^\mu G_{2\mu\kappa}^{(j)bm}(x-y)G_{1\nu}^{(j)c}(x)
		+\partial^\mu G_{1\mu}^{(j)b}(x)G_{2\nu\kappa}^{(j)cm}(x-y) \nonumber \\
		&&-\partial_\nu G_{3\mu\kappa}^{(j)\mu bcm}(x,x,y)
		-\partial_\nu G_{2\mu\kappa}^{(j)bm}(x-y)G_{1}^{(j)\mu c}(x)) 
		-\partial_\nu G_{1\mu}^{(j)b}(x)G_{2\kappa}^{(j)\mu cm}(x-y))
		\nonumber \\
		&&+gf^{abc}\partial^\mu G_{3\mu\nu\kappa}^{(j)bcm}(x,x,y)
		+gf^{abc}\partial^\mu(G_{2\mu\kappa}^{(j)bm}(x-y)G_{1\nu}^{(j)c}(x))
				+gf^{abc}\partial^\mu(G_{1\mu}^{(j)b}(x)G_{1\nu\kappa}^{(j)cm}(x-y))
		\nonumber \\
		&&+g^2f^{abc}f^{cde}(G_{4\mu\nu\kappa}^{(j)\mu bdem}(x,x,x,y)
		+G_{3\mu\nu\kappa}^{(j)bdm}(x,x,y)G_{1}^{(j)\mu e}(x) 
		+G_{2\mu\nu}^{(j)bd}(x,x)G_{2\kappa}^{(j)\mu em}(x-y)\nonumber \\
	&&+G_{3\nu\rho\kappa}^{(j)acm}(x,x,y)G_{1}^{(j)\rho b}(x)
	+G_{2\nu\rho}^{(j)eb}(x,x)G_{2\kappa}^{(j)\rho dm}(x-y) \nonumber \\
	&&+G_{2\nu\rho}^{(j)de}(x,x)G_{2\kappa}^{(j)\rho bm}(x-y)
	+G_{1}^{(j)\mu b}(x)G_{3\mu\nu\kappa}^{(j)dem}(x,x,y)+ \nonumber \\
	&&G_{2\kappa}^{(j)\mu bm}(x-y)G_{1\mu}^{(j)d}(x)G_{1\nu}^{(j)e}(x)+
	G_{1}^{(j)\mu b}(x)G_{2\mu\kappa}^{(j)dm}(x-y)G_{1\nu}^{(j)e}(x)+
	G_{1}^{(j)\mu b}(x)G_{1\mu}^{(j)d}(x)G_{2\nu\kappa}^{(j)em}(x-y)) \nonumber \\
	&&	=gf^{abc}(\partial_\nu K^{(j\varepsilon)bcm}_{3\kappa}(x,x,y)
	+\partial_\nu (\bar P^{(\varepsilon)b}_1(x)K^{(j\varepsilon)cm}_{2\kappa}(x,y))) \nonumber \\
	&&+\partial_\nu (\bar K^{(j\varepsilon)bm}_{2\kappa}(x,y)P^{(\varepsilon)c}_1(x))) 
	+ \delta_{am}g_{\nu\kappa}\delta^4(x-y), \nonumber \\
	 &&\partial^2 P^{(\varepsilon)am}_2(x-y)+gf^{abc}\partial^\mu
	(K^{(\varepsilon,j)bcm}_{3\mu}(x,x,y)+P^{(\varepsilon)bm}_2(x-y)G_{1\mu}^{(j)c}(x)+ \nonumber \\
	&&P^{(\varepsilon)b}_1(x)K_{2\mu}^{(j\varepsilon)cm}(x-y))=\delta_{am}\delta^4(x-y), \nonumber \\
	&&\partial^2 K^{(j\varepsilon)am}_{2\kappa}(x-y)+gf^{abc}\partial^\mu
	(L^{(\varepsilon,j)bcm}_{2\mu\kappa}(x,x,y)+ \nonumber \\
	&&K^{(j\varepsilon)bm}_{2\kappa}(x-y)G_{1\mu}^{(j)c}(x)+P^{(\varepsilon)b}_1(x)G_{2\mu\kappa}^{(j)cm}(x-y))=0. 
\end{eqnarray}
By setting currents to zero and using translation invariance, the above mentioned relations yield
\begin{eqnarray}
\label{eq:ds_3}
    &&\partial^2G_{2\nu\kappa}^{am}(x-y)+gf^{abc}(
		\partial^\mu G_{3\mu\nu\kappa}^{bcm}(0,x-y)+\partial^\mu G_{2\mu\kappa}^{bm}(x-y)G_{1\nu}^{c}(x)
		+\partial^\mu G_{1\mu}^{b}(x)G_{2\nu\kappa}^{cm}(x-y) \nonumber \\
		&&-\partial_\nu G_{3\mu\kappa}^{\mu bcm}(0,x-y)-\partial_\nu G_{2\mu\kappa}^{bm}(x-y)G_{1}^{\mu c}(x)) 
		-\partial_\nu G_{1\mu}^{b}(x)G_{2\kappa}^{\mu cm}(x-y))
		\nonumber \\
		&&+gf^{abc}\partial^\mu G_{3\mu\nu\kappa}^{bcm}(0,x-y)
		+gf^{abc}\partial^\mu(G_{2\mu\kappa}^{bm}(x-y)G_{1\nu}^{c}(x))
				+gf^{abc}\partial^\mu(G_{1\mu}^{b}(x)G_{1\nu\kappa}^{cm}(x-y))
		\nonumber \\
		&&+g^2f^{abc}f^{cde}(G_{4\mu\nu\kappa}^{\mu bdem}(0,0,x-y)
		+G_{3\mu\nu\kappa}^{bdm}(0,x-y)G_{1}^{\mu e}(x) 
		+G_{2\mu\nu}^{bd}(0)G_{2\kappa}^{\mu em}(x-y)\nonumber \\
	&&+G_{3\nu\rho\kappa}^{acm}(0,x-y)G_{1}^{\rho b}(x)
	+G_{2\nu\rho}^{eb}(0)G_{2\kappa}^{\rho dm}(x-y)
	+G_{2\nu\rho}^{de}(0)G_{2\kappa}^{\rho bm}(x-y)
	+G_{1}^{\mu b}(x)G_{3\mu\nu\kappa}^{dem}(0,x-y)+ \nonumber \\
	&&G_{2\kappa}^{\mu bm}(x-y)G_{1\mu}^{d}(x)G_{1\nu}^{e}(x)+
	G_{1}^{\mu b}(x)G_{2\mu\kappa}^{dm}(x-y)G_{1\nu}^{e}(x)+
	G_{1}^{\mu b}(x)G_{1\mu}^{d}(x)G_{2\nu\kappa}^{em}(x-y)) \nonumber \\
	&&	=gf^{abc}(\partial_\nu K^{bcm}_{3\kappa}(0,x-y)+\partial_\nu (\bar P^{b}_1(x)K^{cm}_{2\kappa}(x-y))) 
	+\partial_\nu (\bar K^{bm}_{2\kappa}(x-y)P^{c}_1(x)))
	+ \delta_{am}g_{\nu\kappa}\delta^4(x-y) \nonumber \\
	 &&\partial^2 P^{am}_2(x-y)+gf^{abc}\partial^\mu
	(K^{bcm}_{3\mu}(0,x-y)+P^{bm}_2(x-y)G_{1\mu}^{c}(x)+ \nonumber \\
	&&P^{b}_1(x)K_{2\mu}^{cm}(x-y))=\delta_{am}\delta^4(x-y), \nonumber \\
	&&\partial^2 K^{am}_{2\kappa}(x-y)+gf^{abc}\partial^\mu
	(L^{bcm}_{2\mu\kappa}(0,x-y)+ \nonumber \\
	&&K^{bm}_{2\kappa}(x-y)G_{1\mu}^{c}(x)+P^{b}_1(x)G_{2\mu\kappa}^{cm}(x-y))=0. 
\end{eqnarray}

Finally, let us state the partition function we just obtained that can be written as
\be
Z[j]=\sum_{n=1}^\infty\frac{1}{n!}\int dx_1\ldots dx_nG^{a_1\ldots a_n}_{n\mu_1\ldots\mu_n}(x_1,\ldots,x_n)j^{\mu_1 a_1}(x_1)
\ldots j^{\mu_n a_n}(x_n).
\ee
In the infrared limit, it will be enough to write down for the action
\be
S=\frac{1}{2}\int dx_1dx_2j^{\mu_1 a_1}(x_1)G_{2\mu_1\mu_2}^{a_1a_2}(x_1-x_2)
j^{\mu_2 a_2}(x_2).
\ee

\section{Solution of the Dyson-Schwinger equations and spectrum of the theory}
\label{DS}

The solution of the equations obtained in the preceding section can be obtained by the mapping theorem. So, just stating the 1P- and 2P-correlation functions of the scalar field, we are able to solve the corresponding set of equations for the Yang-Mills theory.

For the 1P-function, one gets for the scalar field \cite{Frasca:2017slg}
\begin{eqnarray}
\phi(x)=\sqrt{\frac{2\mu^4}{m^2+\sqrt{m^4+2Ng^2\mu^4}}}\times
\nonumber \\
{\rm sn}\left(p\cdot x+\chi,\kappa\right),
\end{eqnarray}
being sn a Jacobi elliptical function, $\mu$ and $\chi$ arbitrary integration constants and $m^2=2Ng^2\Delta_\phi(0)$. Then,
\begin{equation}
\kappa=\frac{-m^2+\sqrt{m^4+2Ng^2\mu^4}}{-m^2-\sqrt{m^4+2Ng^2\mu^4}}.
\end{equation}
This is true provided that the following dispersion relation holds
\begin{equation}
    p^2=m^2+\frac{Ng^2\mu^4}{m^2+\sqrt{m^4+2Ng^2\mu^4}}.
\end{equation}
Then, for the Yang-Mills theory, using the mapping theorem in the Landau gauge, one has
\be
A_\mu^a(x)=\eta_\mu^a\phi(x).
\ee
Similarly, for the 2P-function for the scalar field in momenta space, it is \cite{Frasca:2017slg}
\begin{eqnarray}
   \Delta(p)=M{\hat Z}(\mu,m,Ng^2)\frac{2\pi^3}{K^3(\kappa)}\times \nonumber \\
	\sum_{n=0}^\infty(-1)^n\frac{e^{-(n+\frac{1}{2})\pi\frac{K'(\kappa)}{K(\kappa)}}}
	{1-e^{-(2n+1)\frac{K'(\kappa)}{K(\kappa)}\pi}}\times \nonumber \\
	(2n+1)^2\frac{1}{p^2-m_n^2+i\epsilon}
\end{eqnarray}
being
\begin{equation}
M=\sqrt{m^2+\frac{Ng^2\mu^4}{m^2+\sqrt{m^4+2Ng^2\mu^4}}},
\end{equation}
and ${\hat Z}(\mu,m,Ng^2)$ a given constant. This gives rise to a gap equation for the mass shift $m$ on the theory spectrum $m_n$ \cite{Frasca:2017slg}.

\section{Effective potential\label{EF}}

In order to compute the effective potential for the Casimir effect, we consider the simplest geometrical setting. We assume infinite plane boundaries for the axes x,y and periodic boundary condition on the z coordinate. In our case, it is not difficult to write down the expected effective potential having an action made up of a sum of free contributions.

\subsection{Scalar field theory}

The action has the very simple form in Euclidean metric
\begin{equation}
   S=-\frac{1}{2}\int\frac{d^4p}{(2\pi)^4}j(p)\Delta(p)j(-p)=
   \sum_{n=0}^\infty B_n\int\frac{d^4p}{(2\pi)^4}j(p)\frac{1}{p^2+m_n^2}j(-p)
\end{equation}
showing a sum of weighted free propagators. Now, this gives us the effective action
\begin{equation} 
   \Gamma = -\frac{1}{2}\sum_{n=0}^\infty B_n\log{\rm det}(p^2+m_n^2)
\end{equation}
and using the standard relation $\ln{\rm det}={\rm Tr}\log$, with the given boundary conditions one has
\begin{equation}
   -\frac{\Gamma}{T} = E_0 = \frac{1}{2}\sum_{n=0}^\infty B_n\int\frac{dp_0dp_1dp_2}{(2\pi)^3}
   \sum_{n_z=-\infty}^\infty\log\left(p_0^2+p_1^2+p_2^2+\left(\frac{2n_z\pi}{L}\right)^2+m_n^2\right)S
\end{equation}
being $L$ the distance between the two slabs, $S$ the surface of the boundary. In order to get the Casimir contribution we have to evaluate this expression. Firstly, we note that the sum on $n_z$ is well-known in thermal field theory as a Matsubara sum \cite{lebellac}. So, one has
\begin{equation}
   \frac{1}{2}\sum_{n_z=-\infty}^\infty\log\left(\omega^2_n+\left(\frac{2n_z\pi}{L}\right)^2\right)
   =\frac{L\omega_n}{2}+\ln\left(1-e^{-L\omega_n}\right)+constant
\end{equation}
being $\omega_n=\sqrt{p_0^2+p_1^2+p_2^2+m_n^2}$ and the constant is divergent but independent on $L$ or $\omega$. Then, the Casimir term is straightforwardly obtained as \cite{Bordag:2001qi}
\begin{equation}
   E_C(L)=\sum_{n=0}^\infty B_n\int\frac{d^3p}{(2\pi)^3}\ln\left(1-e^{-L\omega_n}\right)S.
\end{equation}
This is the standard result if there is no mass spectrum and no mass gap, that is $m_n\rightarrow 0$. In the infrared limit, being $m_n$ finite, we have to pursue a different approach. We should consider small momenta and a cut-off $\mu$, the same as in the mass spectrum. Then, we are able to evaluate the integral. One has
\begin{equation}
   E_C(L)\approx -\sum_{n=0}^\infty B_ne^{-Lm_n}\frac{\mu^3}{6\pi^2}S\approx -B_0e^{-Lm_0}\frac{\mu^3}{6\pi^2}S
\end{equation}
where the approximation $Lm_0\gg 1$ for the mass gap has been used. We can conclude from this equation that, in the infrared limit, the Casimir contribution is exponentially small. This result should be expected from the simple fact that the appearance of a mass gap makes forces short-ranged.

\subsection{Yang-Mills theory}

Due to the existence of a mapping theorem between scalar and Yang-Mills fields we can draw a similar conclusion for a gluonic field. This conclusion can change when either quarks are present or a mass gap goes to zero increasing momenta scale recovering asymptotic freedom. This makes interesting a study at finite temperature with given boundaries. So, taking at the infrared trivial fixed point the action
\begin{equation}
     S_0=-\frac{1}{2}\int \frac{d^4p}{(2\pi)^4}j^{\mu a}(p)D_{\mu\nu}^{ab}(p)j^{\nu b}(-p)
\end{equation}
being
\begin{equation}
     D_{\mu\nu}^{ab}(x-y)=\delta_{ab}\left(\eta_{\mu\nu}-\frac{p_\mu p_\nu}{p^2}\right)\Delta(p)
\end{equation}
the gluon propagator in the Landau gauge and $\Delta(p)$ the propagator of the scalar field. But if we use current conservation the action is immediately reduced to
\begin{equation}
     S_0=-\frac{1}{2}\int \frac{d^4p}{(2\pi)^4}j^{\mu a}(p)\Delta(p)j_\mu^a(-p)
\end{equation}
that maps perfectly on the action of the scalar field provided a multiplying factor $d(N^2-1)$ that is exactly the number of current products. This means that the result is identical to the one of the scalar field apart an inessential numerical factor and we can conclude that the Casimir effect is exponential small also in Yang-Mills theory as expected.

The result we obtained is consistent with recent lattice computations \cite{Chernodub:2018pmt}. Besides, they get for an estimation of the mass gap of the theory $m_0/\sqrt{\sigma}\approx 1.37$, $\sigma$ is the well-known string tension that in real world is about $(440\ MeV)^2$, while our computations provide $m_0/\sqrt{\sigma}\approx 1.19$ that is in very close agreement with their lattice data.

\section{Conclusions\label{conc}}

Using recent analyses, both theoretical and numerical ones, on the behavior of the propagators of quantum field theories in the infrared limit made quite simple to derive the behavior of the corresponding vacuum in presence of a simple boundary. In the deep infrared we must expect a quite different behavior for a Yang-Mills theory due to a non-trivial vacuum, generally non-perturbatively well-described by a liquid of instantons, due to the appearance of a mass gap in the theory. A massive scalar field in the free case shows a Casimir contribution exponentially small. Here, we can see the same phenomenon to appear as both a massless scalar field theory and Yang-Mills theory acquire mass dynamically displaying a mass gap. But the situation and even more better due to the fact that these theories hit a trivial infrared fixed point making for us very easy to adopt the approaches for free theories. In the near future will be very interesting to extend this analysis to more complex boundaries and include quark matter.

\end{document}